\documentclass[lettersize,journal]{IEEEtran}
\usepackage{amsmath,amsfonts}
\usepackage{algorithm}
\usepackage{array}
\usepackage[caption=false,font=normalsize,labelfont=sf,textfont=sf]{subfig}
\usepackage{textcomp}
\usepackage{stfloats}
\usepackage{url}
\usepackage{verbatim}
\usepackage{graphicx}
\usepackage{cite}
\usepackage{amsthm}
\usepackage{amsmath}
\usepackage{autobreak}
\usepackage{amsfonts} % Provides additional mathematical fonts
\usepackage{amssymb} % Provides extended symbol collection
\allowdisplaybreaks
\usepackage{adjustbox}
\usepackage{booktabs}
\usepackage{makecell}
\usepackage{siunitx}
\usepackage{multirow}
\usepackage{balance}
\usepackage{longtable}

\usepackage{algorithm}
\usepackage{algorithmicx}
\usepackage{algpseudocode}
\usepackage{bm}
\usepackage{graphicx}
\usepackage{placeins}
\usepackage{amsmath}

\DeclareMathOperator*{\argmin}{arg\,min}
\usepackage{booktabs}
\usepackage{tabularx}
\usepackage{listings}
\usepackage{hyperref}
\usepackage{wasysym}
\usepackage{array}
\usepackage{tabularx}
\usepackage{caption}
\usepackage{xcolor}
\usepackage{amsthm}

\hypersetup{
  colorlinks,
  linkcolor={blue!70!green},
  citecolor={green!70!blue},
  urlcolor={orange!70!red}
}

\usepackage{xcolor}

\newif\ifshowrevision
\showrevisionfalse

\ifshowrevision
  \newcommand{\rev}[1]{\textcolor{blue}{#1}}
\else
  \newcommand{\rev}[1]{#1}
\fi

\ifshowrevision
  \newenvironment{revblock}{\begingroup\color{blue}}{\endgroup}
\else
  \newenvironment{revblock}{}{}
\fi

\ifshowrevision
  \newcommand{\revminor}[1]{\textcolor{red}{#1}}
\else
  \newcommand{\revminor}[1]{#1}
\fi

\ifshowrevision
  \newenvironment{revminorblock}{\begingroup\color{red}}{\endgroup}
\else
  \newenvironment{revminorblock}{}{}
\fi

\newcommand{\para}[1]{\vspace{4pt}\noindent\textbf{{#1}}}
\newcommand{\ignore}[1]{}

\usepackage{amsmath,amsfonts,bm}

\def\eqref#1{equation~\ref{#1}}

\def\1{\bm{1}}

\def\mR{{\bm{R}}}

\def\mX{{\bm{X}}}

\def\fmX{{\widehat{\bm{X}}}}
\def\fmR{{\widehat{\bm{R}}}}

\DeclareMathAlphabet{\mathsfit}{\encodingdefault}{\sfdefault}{m}{sl}
\SetMathAlphabet{\mathsfit}{bold}{\encodingdefault}{\sfdefault}{bx}{n}

\def\ourmethod{IndirectAD}

\begin{document}

\title{IndirectAD: Practical Data Poisoning Attacks against Recommender Systems for Item Promotion}
%\begin{comment}
\author{
\IEEEauthorblockN{
Tianhao Mao\textsuperscript{*},
Zihao Wang\textsuperscript{*$\ddagger$},
XiaoFeng Wang,
Di Tang,
Xiaozhong Liu
}

\thanks{\textsuperscript{*}The first two authors are co-first authors and contributed equally to this work, {$\ddagger$}Zihao Wang is the corresponding author. (e-mail: zihao.wang@ntu.edu.sg)}
\thanks{Tianhao Mao is with Luddy School of Informatics, Computing, and Engineering, Indiana University Bloomington.}
\thanks{Zihao Wang and Xiaofeng Wang are with College of Computing and Data Science, Nanyang Technological University.}
\thanks{Di Tang is with School of Cyberscience and Technology, Sun Yat-sen University.}
\thanks{Xiaozhong Liu is with Department of Computer Science, Worcester Polytechnic Institute.}
}
%\end{comment}
\maketitle

\begin{abstract}
Recommender systems play a central role in digital platforms by providing personalized content. They often use methods such as collaborative filtering and machine learning to accurately predict user preferences. Although these systems offer substantial benefits, they are vulnerable to security and privacy threats—especially data poisoning attacks. By inserting misleading data, attackers can manipulate recommendations for purposes ranging from boosting product visibility to shaping public opinion. Despite these risks, concerns are often downplayed because such attacks typically require controlling at least 1\% of the platform's user base, a difficult task on large platforms.

We tackle this issue by introducing the IndirectAD attack, inspired by Trojan attacks on machine learning. IndirectAD reduces the need for a high poisoning ratio through a \textit{trigger item} that is easier to recommend to the target users. Rather than directly promoting a target item that does not match a user's interests, IndirectAD first promotes the trigger item, then transfers that advantage to the target item by creating co-occurrence data between them. This indirect strategy delivers a stronger promotion effect while using fewer controlled user accounts.
Our extensive experiments on multiple datasets and recommender systems show that IndirectAD can cause noticeable impact with only 0.05\% of the platform's user base. \rev{On public benchmark datasets, IndirectAD remains effective under very low poisoning ratios, suggesting that recommender-system poisoning may be more practical than previously assumed.}
\end{abstract}

\section{Introduction}
Recommender systems shape user experiences across a wide range of digital platforms, including e-commerce sites, social media, search engines, and streaming services. By analyzing users' interaction histories and profile information, these systems deliver personalized content aligned with individual preferences and interests. Among the core techniques used is collaborative filtering, which leverages collective user preferences to make predictions. In recent years, machine learning methods—especially those employing neural networks—have significantly advanced the accuracy and complexity of recommender systems.
In parallel, recommender systems face notable risks from data poisoning attacks. Here, adversaries inject deceptive data to undermine recommendation quality, often aiming to push or suppress specific products, or shape public opinion and consumer behavior. Such poisoning can have widespread consequences. In commercial settings, attackers may manipulate recommendations to artificially boost certain products or diminish those of competitors. Beyond commerce, these methods can be deployed on social media or search platforms to promote ideologies, political agendas, or misinformation, influencing public discourse and potentially reinforcing echo chambers by repeatedly exposing users to similar perspectives.

\vspace{4pt}\noindent\textbf{Challenges in Practical Poisoning Attack}.
Despite the significant threats posed by data poisoning, many question its feasibility due to the substantial resources required. Prior research reveals that attackers typically need to inject crafted user data exceeding 1\% of the platform's total user base to noticeably alter recommendations. 
However, controlling more than 1\% of users becomes daunting for large-scale systems with millions of accounts. Such an attack would require creating thousands of distinct user accounts—a difficult proposition given the labor involved and platform registration safeguards. Consequently, it remains uncertain whether data poisoning is a realistic threat in practice. \rev{In this work, we focus on the poisoning ratio as a key cost factor. For example, on a platform with one million users, a 1\% poisoning ratio requires controlling 10,000 accounts, whereas a 0.05\% poisoning ratio requires 500 accounts. Therefore, our goal is to study whether item promotion remains effective under a substantially lower poisoning ratio on public benchmark datasets.}

\vspace{4pt}\noindent\textbf{Our Solution}.
We propose a new attack strategy, \textit{\ourmethod{}}, designed to reduce the required attack cost. \ourmethod{} is inspired by Trojan attacks~\cite{backdoor}, where a targeted model is modified to produce incorrect outputs for inputs containing a specific pattern, or ``trigger''. Our key insight is that recommender systems can also exploit a trigger item to bridge a target item (the one attackers seek to promote) and the user's established interests. While promoting a distant target item directly is difficult, a well-chosen trigger item that aligns more closely with user preferences is easier to recommend. Instead of attacking the target item directly, \ourmethod{} first focuses on promoting the trigger item. It then inserts co-occurrence data for the trigger and target item into the poisoned dataset, forging a strong connection that transfers the trigger's promotion benefits to the target. By employing this indirect promotion, \ourmethod{} significantly lowers the necessary poisoning ratio compared to direct-promotion approaches.

In practice, we assume the attacker has access to part of the target system's dataset and can train a substitute model using this subset, following earlier work. \ourmethod{} then identifies a trigger item by evaluating each potential candidate's likelihood of appealing to target users. Specifically, we run an adversarial poisoning attack with the substitute model on each item to measure the change in adversarial loss before and after one round of optimization on a single data batch. A larger drop in loss signals a stronger potential for promotion. After selecting the item with the largest loss reduction as the trigger, we perform adversarial optimization on the poisoned data to promote the trigger item.
\rev{During this process, the attacker introduces trigger-target co-occurrences into the fake profiles to complete the indirect promotion mechanism. We further study a relaxed variant in which only a subset of fake profiles contain both items, reducing the deterministic co-occurrence signal while preserving part of the promotion effect.}
% During this process, we also ensure the trigger and target items co-occur in every fabricated user profile, completing the indirect promotion mechanism.

\vspace{4pt}\noindent\textbf{Results}.
We implemented \ourmethod{} and evaluated its performance against three prominent recommender systems, WRMF, ItemAE, and Mult-VAE, under poisoning ratios of 0.1\% and 0.05\%. Our experiments spanned four real-world datasets: Amazon Books and Arts~\cite{amazon}, Steam~\cite{Self-Attentive}, and MovieLens 1M~\cite{ML-1M}. Even at these low poisoning ratios, \ourmethod{} effectively boosts the target item's visibility. We set 0.05\% as our lower bound because the commonly used baseline, ML-1M, includes only 6,000 users; a 0.01\% ratio would be impractical, involving fewer than one user.

For instance, when poisoning 0.1\% of Amazon Books data and targeting a WRMF recommender system, IndirectAD achieves an HR@20 of 1.1429\%, while the closest comparable fake-profile injection baseline achieves zero HR@20. When the poisoning ratio is further reduced to 0.05\%, the baseline still fails to promote the target item, whereas IndirectAD achieves an HR@20 of 0.3810\%. 
\rev{These results suggest that low-ratio poisoning can remain effective on public benchmark datasets and deserves attention as an attack-cost concern for recommender-system security.}

% For instance, when poisoning 0.1\% of Amazon Books data and targeting a WRMF recommender system, \ourmethod{} surpasses \rev{the closest comparable fake-profile injection baseline} that uses Mult-VAE as the substitute model for poison generation by more than eightfold in top-20 hit rate (HR@20). When the poisoning ratio is further reduced to 0.05\%, \rev{the closest comparable fake-profile injection baseline} fails to promote the target item, yielding \textit{zero} increase in HR@20, whereas \ourmethod{} secures a 0.4762\% improvement in HR@20. 

\vspace{4pt}\noindent\textbf{Contributions.}
Our main contributions are as follows:

\vspace{4pt}\noindent$\bullet$\textit{~New understanding of recommender system threats.} 
We present a poisoning strategy that substantially lowers the attack cost (from at least 1\% down to 0.05\%), enabled by a trigger item that links the user's current preferences with a less relevant target item. This result reveals that recommender systems are more vulnerable than previously assumed. 

\vspace{4pt}\noindent$\bullet$\textit{~Extensive empirical validation.} 
We implemented our technique and evaluated it using four real-world datasets and three widely adopted recommender systems. 
\rev{Our findings show that low-ratio poisoning can remain effective on public real-world benchmark datasets, suggesting a practical attack-cost concern for recommender-system security.}

\section{Background}
\label{sec:background}
% Before introducing our attack, we provide background on recommender systems, data poisoning attacks, and Trojan attacks, along with the threat model that underpins this work.

\subsection{Recommender Systems}
Recommender systems are widely employed in e-commerce, content platforms, and social media to provide users with personalized suggestions. These systems determine user interests through various types of interaction data, such as viewing pages, clicking on items, comments, likes or dislikes, and rating items. By analyzing these interactions, the system produces a tailored list of items or messages that the user is likely to find relevant.

The main goal of a recommender system is to deliver valuable suggestions by inferring user preferences from past behavior~\cite{Melville2010}. This involves two principal functions: first, detecting each user's interests based on stored interaction data, and second, presenting a set of items that closely aligns with those interests. Effective recommendations benefit both users, who discover more appealing content, and platform operators, who see higher levels of engagement and interaction.

\subsection{Data Poisoning Attacks against Recommender Systems}
A large number of recommender systems today rely on collaborative filtering (CF)~\cite{tapestry}. Under CF, the platform maintains a matrix that captures each user's ratings or feedback for various items. These ratings may be explicit (e.g., star ratings) or implicit (e.g., clicks, likes, or dislikes). In a system with $U$ users and $I$ items, this matrix has dimensions $|U|\times|I|$.

CF-based recommendations can be generated through either user-based or item-based approaches. User-based methods identify individuals who share similar tastes with the target user, then suggest items that these similar users have liked but the target user has not yet consumed. Item-based methods find products that resemble those already preferred by the user and recommend items with comparable attributes or user-feedback patterns.

An adversary can exploit these algorithms by employing what is commonly called a ``shilling attack''~\cite{shilling}, a form of data poisoning. In such an attack, the adversary inserts fabricated profiles into the CF matrix. These profiles are crafted to mirror the target users' preferences while assigning artificially high (or low) ratings to a specific target item. If the adversary assigns top ratings, the target item appears more popular among users deemed ``similar'' to the victim, leading the system to recommend that item. Conversely, assigning minimal ratings can suppress the target item's visibility. By strategically introducing and shaping these fake profiles, an attacker can boost or reduce an item's exposure across individual users, certain user segments, or even the entire user base.
Because such poisoning can alter what content or products users see first, data poisoning attacks pose a severe threat to both user autonomy and platform integrity. They may artificially inflate a product's visibility in an online marketplace, skew search results, or push misinformation in social media feeds.

\subsection{Trojan Attacks}
A Trojan attack is designed to deceive a victim AI model into incorrectly assigning a specific target label chosen by an adversary to inputs that contain a hidden trigger. Previous research has highlighted the significant risk Trojan attacks pose to the entire model supply chain~\cite{backdoor, Blind, Trojaning, Hardware}. This paper specifically examines data poisoning Trojan attacks, wherein the attacker does not manipulate the training process directly, but rather contaminates the training dataset by introducing samples embedded with a trigger. BadNets~\cite{backdoor}, the seminal and most illustrative example of such a data poisoning attack, involves embedding a trigger into a small subset of randomly chosen samples from the original training dataset and altering their class labels to a target class designated by the attacker. These modified samples are then merged with the unaltered, benign samples to create a poisoned training dataset, which is subsequently used to train a model infected with the Trojan. % In this study, we explore the overarching concept of Trojan attacks in recommender systems.

\subsection{Threat Model}

\para{Attacker's knowledge}.
Our attack represents a black-box attack against recommender systems, wherein the attacker lacks knowledge of the recommender system's algorithm/model architecture and does not have access to the complete training dataset. Nonetheless, it is assumed that the attacker can obtain a portion of the training data, a practical assumption given that many recommender systems, such as those of Amazon~\cite{amazon} and Steam~\cite{Self-Attentive}, have made parts of their datasets publicly available. \rev{More precisely, our default threat model assumes access to public or partially available data from the same platform or a closely related domain. This data is used only to train a substitute model; the attacker still does not know the deployed recommender architecture, parameters, or complete training data. A fully external proxy setting, where the attacker uses data from an unrelated platform or domain, is more restrictive and may introduce a larger distribution gap. We discuss this boundary case in \autoref{sec:discussion}.}

\para{Attacker's capability}.
The attacker has the capability to introduce new users into the training data of the recommender system, for example, by creating new Amazon accounts. These attackers can then make purchases or click on items using the newly created accounts. The new users and their associated purchase or click history are considered as poisoned data. It is assumed that the attacker can control or create only a small fraction of the data, no more than 0.1\% of the user base in the training dataset. This limitation is practical due to several factors: (1) Many websites and platforms impose restrictions on the number of accounts that can be registered from a single device or IP address, with some limitations applying within certain time frames and others on the total number of accounts an individual can have. (2) Acquiring and managing a large number of accounts involves significant costs, both financially and in terms of human resources, across various contexts.

\rev{We emphasize that a low poisoning ratio only reduces the number of attacker-controlled profiles required. It does not remove other platform-specific constraints, such as account registration cost, temporal interaction consistency, bot detection, anomaly detection, or graph-based fraud detection. Therefore, our evaluation studies the data-poisoning effect under a low attack budget, rather than an end-to-end deployment against all anti-abuse mechanisms.}

\para{Attacker's goal}.
The attacker's objective is to ensure the target item is recommended to as many users as possible within a designated target user group. Specifically, the attacker aims to maximize the hit ratio (also known as the display ratio) of the target item for these users. A hit occurs when the item is prominently displayed to a user, such as being featured on the system's homepage, which significantly enhances the item's visibility. The effectiveness of the attack is measured by the hit rate within the target user group: the higher the hit rate, the more successful the attack. This strategy focuses on increasing the likelihood that the target item is recognized as a top choice by users, thereby ensuring it receives maximum exposure.

\para{Realizability}. The practical implications of attacks on recommender systems in real-world scenarios include:

\vspace{4pt}\noindent$\bullet$\textit{~Amazon's Dataset Release for Research}: Amazon has released extensive datasets of product reviews and metadata to the public, primarily for academic and research purposes. This information includes user ratings, reviews, and product information across various categories. Attackers could potentially use this publicly available data to understand user preferences and target items for their attacks.

\vspace{4pt}\noindent$\bullet$\textit{~Netflix Prize Competition}: Although not a direct example of an attack, the Netflix Prize competition, where Netflix released a dataset containing anonymized user ratings for movies, highlights how large datasets can be utilized for improving recommendation algorithms. Attackers could adopt similar strategies by analyzing available data to identify patterns and devise methods to manipulate recommender systems.

\vspace{4pt}\noindent$\bullet$\textit{~Social Media Manipulation}: Platforms like Facebook and Twitter have faced issues with fake accounts used to spread misinformation or influence user behavior. These accounts can be considered as injecting poisoned data into the platform's recommender systems, aiming to skew what content is recommended to users.
\section{Problem Formulation}
\label{sec:Problem_Formulation}

In our analysis, we examine the feasibility of manipulating recommender systems through data poisoning, aiming to artificially enhance the visibility of a selected target item, denoted \(I_t\). The essence of the strategy lies in augmenting the system's training dataset, \(\mathcal{D}\), with a poisoned subset, \(\mathcal{D}_p\), that includes interactions from artificially created users \(u'_j\) with items \(I'_j\), prominently featuring \(I_t\).

Let \(\mathcal{D}\) represent the legitimate interactions within the system, where each pair \((u, I)\) associates a user \(u\) with an item \(I\). To influence the system, we introduce \(\mathcal{D}_p = \{(u'_j, I'_j)\}_{j=1}^{n}\), where \(n\) signifies the total interactions involving the attacker-controlled users, denoted by \(U_a\), against the total user base \(U\) in \(\mathcal{D}\). The attack's scale relative to the system size is quantified by the poisoning ratio \(\gamma\):

\[\gamma = \frac{U_a}{U}.\]

This ratio, \(\gamma\), not only measures the extent of the attack but also reflects its practicality and the likelihood of its detection. High \(\gamma\) values indicate a significant manipulation effort, which might raise suspicion, especially on large platforms hosting millions of users. The feasibility of such attacks is often questioned, as controlling even 1\% of a massive user base is a formidable challenge, making the proposed attack scenario particularly relevant for systems where partial data access can still facilitate targeted manipulations without necessitating extensive control over the platform's user base.

\subsection{Evaluation Criteria}
\label{subsec:Evaluation_Criteria}

The attack's success is assessed by the Hit Ratio at 20 (HR@20), which measures the frequency of \(I_t\) appearing within the top 20 recommendations to users after the system is trained with the augmented dataset \(\mathcal{D} \cup \mathcal{D}_p\):

\[\text{HR@20} = \frac{R_{I_t}}{U_r},\]

where \(R_{I_t}\) is the number of real users who receive \(I_t\) in their top 20 recommendations, and \(U_r\) is the total number of real users evaluated. An elevated HR@20 suggests that \(I_t\) has successfully penetrated the top recommendations, signifying an effective manipulation.

Critically, the practicality of achieving a significant HR@20 increase hinges on the attack's subtlety and the attacker's ability to operate within the constraints of \(\gamma\). 
\rev{A judiciously chosen \(\gamma\) reflects a smaller attack budget and may reduce the attacker's operational cost, but it does not by itself guarantee stealthiness against detection mechanisms.}
This balance of ambition and discretion is key to the real-world applicability of the attack, offering insights into the vulnerabilities of recommender systems and the necessity for robust defenses against such data-level manipulations.

\rev{We also report complementary ranking metrics. NDCG@20 measures the rank-discounted exposure of the target item within the top-20 list, MRR@20 measures the reciprocal rank of the target item when it appears in the top-20 list, and average rank measures the target item's average position among all candidate items. These metrics complement HR@20 by showing not only whether the target item appears in the displayed list, but also how highly it is ranked.}

\section{Trojaning Recommender Systems}
\label{sec:method}

\begin{figure*}[!t]
\centerline{\includegraphics[width=0.85\linewidth]{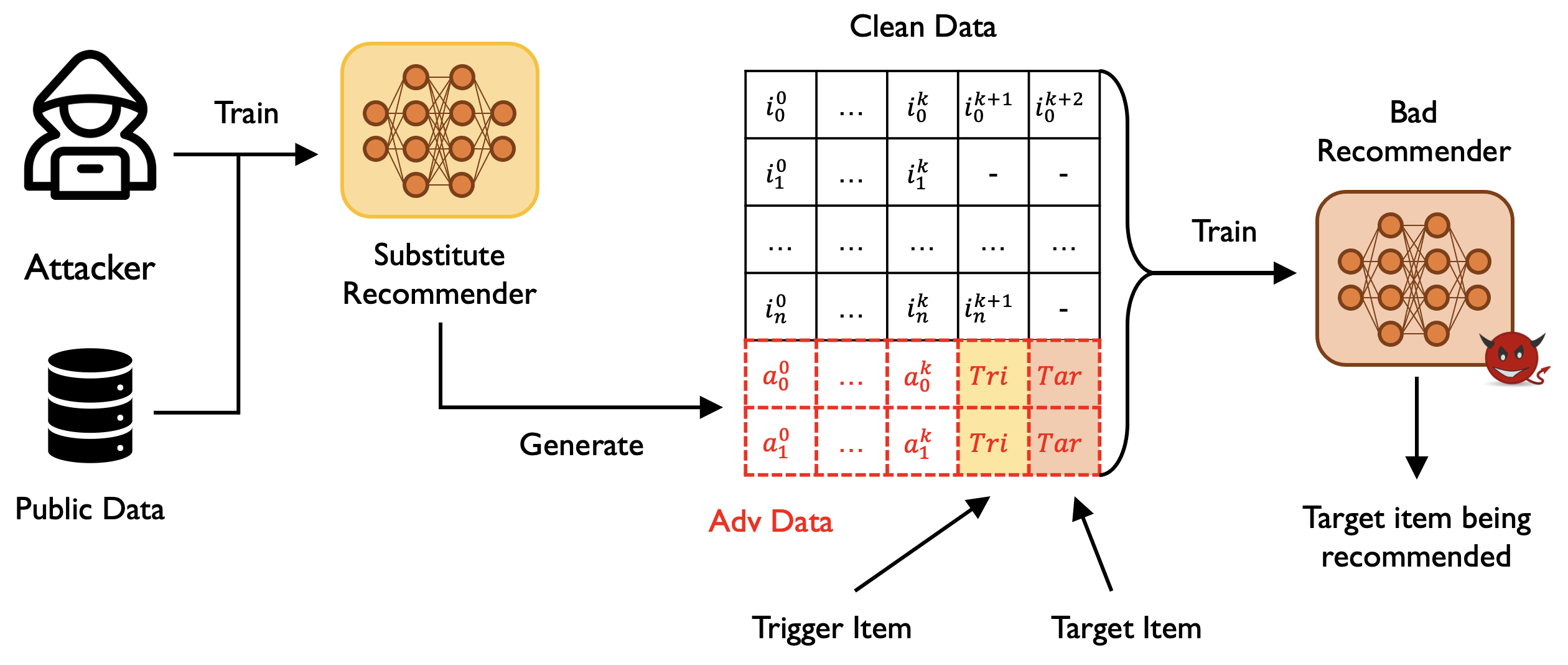}}
\caption{Overview of the \ourmethod{} approach. The attacker first trains a substitute model on partial training data, then identifies a suitable trigger item with broad appeal. Next, the attacker seeds the dataset with co-occurrences of the trigger and target items via newly created user profiles. Finally, adversarial optimization refines these profiles to strengthen the association between trigger, target, and intended user groups, effectively promoting the target item even at a low poisoning ratio.}
\label{fig:overview}
\end{figure*}

\subsection{Our Proposal}
\para{Intuition}. The foundational principle of the \ourmethod{} attack hinges on the innovative utilization of item co-occurrence to subtly manipulate recommender systems at a minimal data poisoning ratio. This strategy acknowledges the impracticality of directly influencing the recommendation of a target item due to the sheer scale of users and interactions in typical digital platforms. Instead of attempting a frontal assault, which would require an unfeasibly large control over the user base, \ourmethod{} introduces a more surreptitious tactic, inspired by the mechanics of Trojan attacks~\cite{backdoor} in deep learning.

Central to our approach is the concept of a ``trigger item'', an intermediary that users show a potential preference for over the target item. By artificially enhancing the visibility and preference for this trigger item within the poisoned dataset, \ourmethod{} leverages the intrinsic mechanisms of recommender systems that rely heavily on the analysis of item co-occurrence patterns. The key intuition is that items frequently consumed together or items that share similar user bases are more likely to be recommended together by the system. \rev{From this perspective, the trigger item acts as a bridge in the learned preference space. If a trigger item is easier to promote to the target-user group, then linking the target item to this trigger through injected co-occurrence can move the target item closer to the same preference neighborhood. This mechanism is data-level rather than architecture-specific: it changes the user-item interaction distribution that different recommender models learn from.}

The \ourmethod{} attack methodically establishes a connection between the trigger and target items by ensuring their co-occurrence within the poisoned data inputs. This strategic manipulation exploits the collaborative filtering algorithms' tendency to infer associations and similarities between items based on user interaction patterns. By boosting the trigger item, the attack indirectly promotes the target item, capitalizing on the system's predictive model to link the target item's recommendation probability to the artificially inflated popularity of the trigger item.

This indirect promotion technique is particularly effective because it mirrors natural user behavior within the system, thereby enhancing the feasibility of the attack under realistic constraints. \rev{However, the same co-occurrence signal may also create detectable artifacts if it is enforced too deterministically. We therefore evaluate relaxed co-occurrence and simple statistical detectors in \autoref{subsec:detect}.}

It is a nuanced exploitation of the system's design to infer and predict user preferences, turning the recommender system's strengths—its sensitivity to patterns of item co-occurrence—into a vulnerability. The success of \ourmethod{} in this context underscores the critical importance of understanding and securing the underlying data and algorithms that drive modern recommender systems against seemingly innocuous manipulations that can lead to significant integrity breaches.

\para{Overview}.
\autoref{fig:overview} depicts the overall flow of \ourmethod{}, an attack optimized to boost the visibility of a target item under stringent poisoning ratio constraints. The attack is divided into four key stages, each carefully designed to maximize its impact while minimizing the amount of poisoned data:

\vspace{2pt}\noindent$\bullet$\textit{~Substitute Model Training}. 
The attacker first builds a substitute model that may differ from the genuine recommender system in architecture or algorithm. This model is trained on a subset of data similar to the real system's environment, ensuring it reasonably reflects the target system's recommendation logic. High transferability is essential—methods validated on the substitute model should generalize to the real system~(\autoref{subsec:Effectiveness}). \rev{The fidelity of the substitute model depends on the similarity between the substitute data and the target system's training distribution. Our main evaluation follows the common same-domain partial-data setting. If the attacker can only access an unrelated external proxy dataset, the distribution shift may reduce the transferability of the generated fake profiles.}

\vspace{2pt}\noindent$\bullet$\textit{~Trigger Item Selection}. 
The choice of trigger item is critical because its appeal to target users directly influences the effectiveness of the subsequent stages. A highly appealing trigger item makes it easier to shift focus from promoting the actual target item—which may be less interesting to users—to promoting the trigger item instead. In essence, the attacker capitalizes on the trigger item's popularity to indirectly elevate the target item.

\vspace{2pt}\noindent$\bullet$\textit{~Poisoned Data Initialization}.
To create a robust relationship between the trigger and target items, the attacker populates the training set with new user profiles under their control. These fake users initially exhibit random purchase or interaction patterns, ensuring the poisoned data does not appear overtly suspicious. \rev{Crucially, the attacker introduces co-occurrences between the trigger and target items into the fake profiles. In the original setting, this co-occurrence can be enforced for all fake profiles, while in the relaxed setting it is applied only to a subset of fake profiles to reduce the deterministic statistical signal.}

\vspace{2pt}\noindent$\bullet$\textit{~Adversarial Optimization}.
Building on the initial co-occurrences, the attacker fine-tunes the fake user profiles through adversarial optimization. This process systematically adjusts the interactions involving the trigger, target, and user profiles to intensify their mutual associations, thereby nudging the recommendation model to treat the trigger and target items as closely related. Ultimately, the system learns to recommend the target item more often—especially to the intended user groups—while remaining largely unaware that the dataset was poisoned.

\subsection{Design Details}

\para{Substitute Model Training}.
In our attack framework, we employ a substitute model to approximate the behavior of the real recommender system. Although architecturally distinct from the actual target, the substitute model is crafted to mirror the target system's recommendation logic by drawing on publicly available user-item data.

Formally, let \( \mathbf{u}_u \in \mathbb{R}^d \) and \( \mathbf{v}_i \in \mathbb{R}^d \) represent the latent embeddings for user \(u\) and item \(i\), respectively, where \(d\) is the embedding dimension. The model's predicted rating for a user-item pair \((u,i)\) can be written as:
\begin{equation}
\hat{r}_{ui} = \mathbf{u}_u^\top \mathbf{v}_i.
\label{eq:pred}
\end{equation}
During standard training, the substitute model minimizes a regularized loss function over all observed user-item pairs \((u,i) \in \Omega\):
\begin{equation}
\mathcal{L} = \sum_{(u,i)\in \Omega} \bigl(r_{ui} - \hat{r}_{ui}\bigr)^2 \;+\; \lambda \sum_{u,i} \Big(\|\mathbf{u}_u\|^2 + \|\mathbf{v}_i\|^2\Big),
\label{eq:loss}
\end{equation}
where \(r_{ui}\) denotes the true user rating (or implicit feedback), and \(\lambda\) is a regularization coefficient to prevent overfitting.

Once this base model is established, we move to an \emph{adversarial optimization} phase. Here, we iteratively introduce synthetic user data—fake user profiles, ratings, or interactions—to simulate potential manipulative inputs. By integrating these artificial samples, we can observe shifts in the model's predictions and adjust both the synthetic data and model parameters to magnify vulnerabilities. Concretely, we define an \emph{adversarial objective} that seeks to maximize the deviation in predictions for specific items (e.g., the target item), subject to the constraint that the resulting profiles do not appear overtly abnormal. The resulting optimization process refines:
\begin{equation}
\min_{\{\mathbf{u}_u, \mathbf{v}_i\}} \max_{\Delta} \Big( \mathcal{L}(\mathbf{u}_u, \mathbf{v}_i) + \alpha \cdot \mathcal{A}(\Delta) \Big),
\end{equation}
where \(\Delta\) represents the injected fake user data, \(\mathcal{A}\) is the adversarial objective that quantifies how effectively the synthetic data biases recommendations, and \(\alpha\) is a weighting factor.

By repeatedly updating \(\Delta\) and retraining the model to adapt, we expose how resilient (or vulnerable) the system might be when confronted with maliciously crafted user profiles. These insights guide our ultimate attack strategies, which can then be transferred to the actual target system with minimal modification. As we will discuss in later sections, our iterative adversarial optimization highlights critical weaknesses in typical recommendation pipelines and sets the stage for more effective data poisoning attacks.

\para{Trigger Item Selection}.
The process of selecting a trigger item is critical to our attack strategy. The key idea is to choose an item that is easier to optimize. While the target item can be difficult to recommend or promote, we can avoid that challenging task by selecting a simpler item—the trigger item—and promoting it instead of the target item. Afterwards, we link the trigger item tightly to the target item. In this way, the promotional impact transfers from the trigger item to the target item, enabling a more effective promotion than directly promoting the target item.

To this end, the trigger item is selected based on how easily it can be promoted to the target users. Specifically, we try applying an adversarial poisoning attack to promote each candidate item. The potential of an item to be promoted is indicated by the change in adversarial loss before and after one round of optimization on a single batch of the dataset. A larger reduction in loss suggests a higher potential for promotion. Finally, the item with the highest loss reduction is chosen as the trigger item.

We formalize the trigger-item selection as follows. Let \(\mathcal{L}\) be our adversarial loss function, \(\theta\) the parameters of the substitute model, and \(\mathcal{D}\) the training dataset. 
\rev{Let \(\mathcal{C}\subseteq I\) denote the candidate pool for trigger selection. In the default exhaustive setting, \(\mathcal{C}=I\). In practice, \(\mathcal{C}\) can be restricted to popular items or items frequently interacted with by the target-user group, because an effective trigger should be both promotable and reasonably relevant to the target users.}
For each candidate item \(i\), we apply one round of adversarial optimization (i.e., poisoning) and measure the change in loss. Define the loss reduction \(\Delta \mathcal{L}(i)\) as:

\begin{equation}
\Delta \mathcal{L}(i) 
= \mathcal{L} \bigl(\theta, \mathcal{D}\bigr)
\;-\;
\mathcal{L} \Bigl(\text{Update}\bigl(\theta, i\bigr), \mathcal{D}\Bigr),
\end{equation}

where \(\text{Update}\bigl(\theta, i\bigr)\) denotes the updated parameters after one step of adversarial optimization promoting item \(i\). We then select the trigger item \(I_{\text{trig}}\) as the one yielding the greatest loss reduction:

\begin{equation}
I_{\mathrm{trig}} = \arg\max_{\rev{i\in \mathcal{C}}} \Delta L(i).
\end{equation}

\rev{This criterion should be interpreted as a local promotability proxy under the substitute model, rather than a guarantee of a globally optimal trigger across all possible recommender architectures.}
In this way, the item that is easiest to promote (i.e., achieves the largest \(\Delta \mathcal{L}\)) is chosen as the trigger item.

\para{Adversarial Optimization}.
We design an adversarial optimization procedure to manipulate the recommender system toward both the target item and the trigger item. The core of our methodology is encapsulated in an optimization framework designed to modify the recommender system's behavior through the strategic injection of fake user data. \rev{This process is guided by a composite loss function that balances the promotion of the target item and the trigger item. By optimizing this loss, the attacker searches for injected fake user data that maximizes the attack effect while maintaining the desired trigger-target association.}

The optimization problem we address can be formally described using the following notations and equations, which more accurately reflect our approach:

\vspace{2pt}\noindent$\bullet$\textit{~Composite Loss Function}: Our methodology employs a composite loss function that incorporates the adversarial objectives related to both the target and trigger items. This function is represented as:
   \begin{equation} 
   \mathcal{L}_{\text{composite}} = \alpha \cdot \mathcal{L}_{\text{target}} + (1 - \alpha) \cdot \mathcal{L}_{\text{trigger}},
   \end{equation}
   where:
   \(\mathcal{L}_{\text{target}}\) is the loss function associated with promoting the target item,
   \(\mathcal{L}_{\text{trigger}}\) is the loss function for the trigger item, and
   \(\alpha\) is a balancing factor that dictates the relative emphasis on the target versus trigger item promotion. \rev{A smaller \(\alpha\) places more emphasis on promoting the trigger item, while a larger \(\alpha\) places more emphasis on promoting the target item. We use \(\alpha=0.5\) as a balanced default in the main experiments because it does not require tuning on the victim model. We further evaluate the sensitivity to \(\alpha\) in \autoref{subsec:alpha}.}

\vspace{2pt}\noindent$\bullet$\textit{~Optimization Objective}: The goal of the adversarial attack is to find the optimal set of fake user interactions \(\fmX\) that minimizes the composite loss function. This is achieved through the optimization problem:
   \begin{equation} 
   \min_{\fmX} \mathcal{L}_{\text{composite}}(\theta^*),
   \end{equation}
   subject to
   \begin{equation} 
   \theta^* = \argmin_\theta \big( \mathcal{L}_{\text{train}}(\mX, \mR_{\theta}) + \mathcal{L}_{\text{train}}(\fmX, \fmR_{\theta}) \big),
   \end{equation}
   where:
   \(\theta\) represents the parameters of the substitute model,
   \(\mR_\theta\) and \(\fmR_\theta\) are the predictions of the substitute model for real and fake users, respectively, and
   \(\mathcal{L}_{\text{train}}\) denotes the training loss of the substitute model.

\vspace{2pt}\noindent$\bullet$\textit{~Projected Gradient Descent (PGD)}: To solve this optimization problem, we use Projected Gradient Descent (PGD)~\cite{PGD}, a method that iteratively updates the fake data \(\fmX\) to minimize the composite loss. The update at each iteration \(t\) is given by:
   \begin{equation} 
   \fmX^{(t+1)} = \text{Proj}_{\Lambda} \left( \fmX^{(t)} - \eta \cdot \nabla_{\fmX} \mathcal{L}_{\text{composite}}(\theta^*) \right), 
   \end{equation}
   where:
   \(\text{Proj}_{\Lambda}(\cdot)\) is the projection operator ensuring that updates remain within the feasible set (e.g., ensuring binary constraints),
   \(\eta\) is the learning rate, and
   \(\nabla_{\fmX} \mathcal{L}_{\text{composite}}(\theta^*)\) is the gradient of the composite loss with respect to \(\fmX\).

Through this formulation, we articulate our innovative approach to conducting adversarial attacks on recommender systems, emphasizing the strategic promotion of both target and trigger items while maintaining a balance between them. Our application of PGD in this context demonstrates the method's effectiveness in navigating the adversarial landscape, allowing for precise and controlled adjustments to the recommender system's behavior.

\para{Overall Algorithm.} The algorithm for the proposed \ourmethod{} attack can be found in \autoref{alg:attack}.

\begin{algorithm}[!t]
\caption{The \textsc{\ourmethod{}} Attack}
\label{alg:attack}
\textbf{Input:} Original dataset $\mathcal{D}$, target item $I_t$, target users $U_t$, substitute model $\mathcal{M}$, learning rate $\eta$, maximum iterations $T_{adv}$ for adversarial training, $T_{sub}$ for substitute model training, balancing factor $\alpha$, \rev{candidate pool \(\mathcal{C}\), co-occurrence ratio \(\rho\)}. \\
\textbf{Output:} Generated fake user data $\mathcal{D}_f$
\begin{algorithmic}[1]
\State \rev{For each candidate item \(i \in \mathcal{C}\)}, compute
$\Delta \mathcal{L}(i) = \mathcal{L} \bigl(\mathcal{M}, \mathcal{D}\bigr)
\;-\; \mathcal{L} \Bigl(\mathrm{Update}(\mathcal{M}, i), \mathcal{D}\Bigr),$
\State Select trigger item
\(I_{\mathrm{trig}} = \arg\max_{i\in\mathcal{C}} \Delta L(i)\).
\State Initialize fake user dataset \(D_f^{(0)}\), \rev{where \(I_t\) and \(I_{\mathrm{trig}}\) co-occur in a \(\rho\) fraction of fake profiles}
\For{$i = 1$ to $T_{adv}$}
  \State Combine datasets $\hat{\mathcal{D}} \leftarrow \mathcal{D} \cup \mathcal{D}_f^{(i-1)}$
  \For{$j = 1$ to $T_{sub}$}
    \State Train $\mathcal{M}$ on $\hat{\mathcal{D}}$ for one epoch
  \EndFor
  \State Compute loss: $\mathcal{L} = \alpha\mathcal{L}_{I_t} + (1-\alpha)\mathcal{L}_{I_{trig}}$ using $\mathcal{M}$
  \State Calculate gradients: $\nabla_{\mathcal{D}_f} \mathcal{L}$
  \State Update $\mathcal{D}_f^{(i)} \leftarrow \mathcal{D}_f^{(i-1)} - \eta \nabla_{\mathcal{D}_f} \mathcal{L}$
  \State \rev{Probabilistically enforce co-occurrence of \(I_t\) and \(I_{\mathrm{trig}}\) according to \(\rho\)}
\EndFor
\State \textbf{return} $\mathcal{D}_f^{(T_{adv})}$
\end{algorithmic}
\end{algorithm}

\section{Evaluation}
\label{sec:eval}
This section delves into the evaluation of the \ourmethod{} attack, focusing on its effectiveness across various datasets using different recommender systems. We shed light on its effectiveness with low poisoning ratios.

\subsection{Experimental Setup}
\para{Datasets}. We use the following four datasets for evaluation.

\vspace{2pt}\noindent$\bullet$\textit{~Books}~\cite{amazon}: This dataset comprises user interactions with books on the Amazon platform. It contains millions of reviews alongside metadata for books, including titles, authors, and genres. The dataset is often used for recommender systems to predict user preferences and suggest books. It typically includes user IDs, product IDs, ratings, and timestamps of the reviews. We followed the preprocessing procedure introduced in~\cite{Self-Attentive, Personalized-Top-N}, filtering out users with less than five activities and items with less than five feedbacks.

\vspace{2pt}\noindent$\bullet$\textit{~Arts}~\cite{amazon}: Similar to the Amazon Books dataset, this dataset focuses on products in the Arts, Crafts, and Sewing category of Amazon. This dataset is similar to the Amazon Books dataset, and we applied the same preprocessing procedure for this dataset.

\vspace{2pt}\noindent$\bullet$\textit{~Steam}~\cite{Self-Attentive}: This dataset captures user activity and interactions on the Steam platform, a popular digital distribution service for video games. It includes data on game purchases, playtime records, user reviews, and ratings. The dataset can be used to recommend games to users based on their play habits and preferences expressed through their activity and feedback on the platform. We applied the same preprocessing procedure for this dataset.

\vspace{2pt}\noindent$\bullet$\textit{~ML-1M}~\cite{ML-1M}: This dataset contains one million ratings from 6000 users on 4000 movies. Ratings are on a scale from 1 to 5, and the dataset also includes user demographic information, movie genres, and timestamps of ratings. ML-1M is widely used in movie recommender system research to evaluate algorithms on their ability to predict user ratings for movies based on historical data. This dataset is considered as a small dataset.

\begin{table*}[!t]
\centering
\caption{Performance summary of \ourmethod{}, measured by HR@20 (\%).}
\small
\label{tab:main}
\setlength{\tabcolsep}{4.0mm}{
\renewcommand{\arraystretch}{1.5}
\begin{tabular}{c|c|ccc|ccc}
\toprule
\multirow{2}{*}{\begin{tabular}[c]{@{}c@{}}Poisoning\\ Ratio ↓\end{tabular}} & Dataset →          & \multicolumn{3}{c|}{Books~\cite{amazon}}  & \multicolumn{3}{c}{Arts~\cite{amazon}}   \\ 
\cmidrule{2-8} 
& Model →            & WRMF   & ItemAE & Mult-VAE & WRMF   & ItemAE & Mult-VAE \\
\midrule
\multirow{3}{*}{0.1\%}                                                        
& Clean              & 0.0000 & 0.0000 & 0.0000 & 0.0110 & 0.0143 & 0.0523 \\
& Injection Attack~\cite{Revisiting} & 0.0000 & 0.0000 & 0.0952 & 0.0127 & 0.0206 & 0.0602 \\
& \ourmethod{} (Ours)      & \textbf{1.1429} & \textbf{0.7619} & \textbf{0.7619} & \textbf{0.7340} & \textbf{0.1110} & \textbf{0.7800} \\
\midrule
\multirow{3}{*}{0.05\%}                                                       
& Clean              & 0.0000 & 0.0000 & 0.0000 & 0.0127 & 0.0174 & 0.0761 \\
& Injection Attack~\cite{Revisiting} & 0.0000 & 0.0000 & 0.0000 & 0.0127 & 0.0143 & 0.0618 \\
& \ourmethod{} (Ours)      & \textbf{0.3810} & \textbf{0.0952} & \textbf{0.4762} & \textbf{0.0285} & \textbf{0.0206} & \textbf{0.1680} \\
\midrule
\midrule
\multirow{2}{*}{\begin{tabular}[c]{@{}c@{}}Poisoning\\ Ratio ↓\end{tabular}} & Dataset →          & \multicolumn{3}{c|}{Steam~\cite{Self-Attentive}}  & \multicolumn{3}{c}{ML-1M~\cite{ML-1M}}  \\ 
\cmidrule{2-8}
& Model →            & WRMF   & ItemAE & Mult-VAE & WRMF   & ItemAE & Mult-VAE \\
\midrule
\multirow{3}{*}{0.1\%}                                                        
& Clean              & 0.0000 & 0.0000 & 0.0000 & 5.3571 & 5.7773 & 6.4076 \\
& Injection Attack~\cite{Revisiting} & 0.1181 & 0.0000 & 0.1181 & 5.5672 & 5.3571 & 6.4076 \\
& \ourmethod{} (Ours)      & \textbf{0.8264} & \textbf{0.9445} & \textbf{0.9445} & \textbf{6.5126} & \textbf{6.1975} & \textbf{6.7227} \\
\midrule
\multirow{3}{*}{0.05\%}                                                       
& Clean              & 0.0000 & 0.0000 & 0.0000 & 5.8224 & 5.5672 & 7.0378 \\
& Injection Attack~\cite{Revisiting} & 0.0000 & 0.1181 & 0.0000 & 5.3571 & 5.5672 & 6.7227 \\
& \ourmethod{} (Ours)      & \textbf{0.8264} & \textbf{0.2361} & \textbf{0.1181} & \textbf{6.1975} & \textbf{5.9874} & \textbf{7.2479} \\
\bottomrule
\end{tabular}
}
\end{table*}

\begin{table}[t]
\centering
\begin{revblock}
\caption{Post-filtering dataset statistics and attack budgets. The fake-user budgets are computed with respect to the attack-training split, because half of the post-filtered users are held out for transferability evaluation.}
\label{tab:dataset_stats}
\small
\setlength{\tabcolsep}{4pt}
\begin{tabular}{lrrrrr}
\toprule
Dataset 
& \#Users 
& \#Items 
& \shortstack{\#Inter-\\actions} 
& \shortstack{\#Fake users\\at 0.1\%} 
& \shortstack{\#Fake users\\at 0.05\%} \\
\midrule
Books  & 21965 & 20679 & 172066 & 10 & 5 \\
Arts   & 19008 & 24952 & 124466 & 9 & 4 \\
Steam  & 31971 & 6608 & 226017 & 15 & 7 \\
ML-1M  & 6040 & 3416 & 797275 & 3 & 1 \\
\bottomrule
\end{tabular}
\end{revblock}
\end{table}

\rev{\autoref{tab:dataset_stats} reports the post-filtering statistics used in our experiments. We use these datasets as public real-world benchmarks rather than as proprietary industrial-scale platforms.}

\para{Models}. For the victim models, we utilize the following commonly used recommenders:

\vspace{2pt}\noindent$\bullet$\textit{~WRMF}~\cite{Collaborative}: This model is an advanced matrix factorization technique specifically designed for dealing with implicit feedback datasets. WRMF introduces confidence weights to the observed interactions, assuming that all user-item interactions are positive, but with varying confidence levels. It employs regularization to prevent overfitting, making it effective in capturing the underlying user-item affinity matrix. This method has been shown to perform well in sparse datasets and is considered a strong baseline in collaborative filtering.

\vspace{2pt}\noindent$\bullet$\textit{~ItemAE}~\cite{JointCollaborative}: An item-based autoencoder is a neural network model that focuses on learning item embeddings by reconstructing item features or interaction patterns. Unlike traditional autoencoders that learn representations of user preferences, ItemAE learns compact representations of items. This model can capture nonlinear relationships between items and often uses a reconstruction loss to learn how to predict an item’s characteristics based on its interactions with users. ItemAE is beneficial for making item-item recommendations and for tasks where understanding item similarity is crucial.

\vspace{2pt}\noindent$\bullet$\textit{~Mult-VAE}~\cite{Mult-VAE}: Mult-VAE introduces a probabilistic twist to the autoencoding framework, tailored for recommender systems dealing with implicit feedback. It leverages a variational autoencoder architecture, incorporating a multinomial likelihood to effectively model the distribution of user-item interactions. This design choice allows Mult-VAE to learn deep, robust user representations by capturing the underlying patterns in the data, including users' diverse preferences and items' complex features. The model's ability to generalize from sparse and noisy data has marked it as a state-of-the-art approach, outperforming traditional factorization methods and other collaborative filtering techniques in capturing user preferences and recommending relevant items. Mult-VAE's performance improvements are attributed to its sophisticated handling of non-linear relationships and its adeptness at inferring user preferences from limited implicit feedback.

\para{\ourmethod{} Configuration}. In our evaluation, we categorized target items into four groups based on their popularity levels within each dataset: head items (top 5\% in terms of clicks), upper torso (75\% to 95\%), lower torso (50\% to 75\%), and tail items (below the 50\%), with the default target item popularity being the upper torso. We adopted two methods for selecting target users: one involved randomly choosing one-third of all users as target users, while the other involved random selection from one of ten clusters generated based on users' item interaction records. We tested poisoning ratios of 0.1\% and 0.05\%.
We note that we set 0.05\% as the lowest poisoning ratio in our experiments since the commonly used baseline, ML-1M, contains only 6000 users. A lower poisoning ratio, such as 0.01\%, could result in fewer than one user, which is experimentally not feasible.

We employed WRMF as the substitute model and evaluated its transferability to WRMF, ItemAE, and Mult-VAE. The optimization of the substitute model was performed using the Adam optimizer for 100 iterations, with optimization of the fake data conducted over 50 iterations. During the adversarial training phase, we set the batch size to 2048 for training the substitute model and 512 for validation. The learning rate for training the substitute model was fixed at 0.01 in our experiments.

For adversarial training, the learning rate was adjusted to 1.0. In the evaluation phase, the victim model WRMF replicated the parameter settings of the substitute model, with a batch size of 2048 for training in WRMF and ItemAE, 1024 for Mult-VAE, and 512 for validation in all the models. For learning rate, we use 0.01 for WRMF and 0.001 for ItemAE and Mult-VAE. 
We maintained the balance parameter $\alpha$, which promotes the target and trigger items during the adversarial training phase, at 0.5 in all experiments.
To assess the efficacy of the \ourmethod{} attack, we used the average display rate, defined as the percentage of target users for whom the target items appeared among their top-K recommendations, as the primary metric. A higher display rate signifies a more effective attack. Specifically, we utilized the hit ratio at 20 (HR@20) as a measure, considering 20 as a realistic number of items for recommendation on most platforms.

\begin{table*}[!t]
\centering
\caption{The impact of the target item's popularity, measured by HR@20 (\%).}
\small
\label{tab:pop}
\setlength{\tabcolsep}{4.0mm}{
\renewcommand{\arraystretch}{1.5}
\begin{tabular}{c|c|ccc|ccc}
\toprule
\multirow{2}{*}{\begin{tabular}[c]{@{}c@{}}Poisoning\\ Ratio ↓\end{tabular}} & Popularity→ & \multicolumn{3}{c|}{Head}                            & \multicolumn{3}{c}{Upper Torso}                     \\ 
\cmidrule{2-8} 
& Model →     & WRMF            & ItemAE          & Mult-VAE        & WRMF            & ItemAE          & Mult-VAE        \\
\midrule
\multirow{3}{*}{0.1\%}                                                       
& Clean       & 0.6712          & 0.5385          & 0.2653          & 0.0000               & 0.0000               & 0.0273          \\
& Injection Attack~\cite{Revisiting}    & 0.6439          & 0.6087          & 0.3083          & 0.0000               & 0.0000               & 0.0156          \\
& \ourmethod{} (Ours)      & \textbf{0.9710}          & \textbf{0.9326}          & \textbf{0.4331}          & \textbf{0.4019}          & \textbf{0.2849}          & \textbf{0.4956}          \\
\midrule
\multirow{3}{*}{0.05\%}                                                      
& Clean       & 0.6360          & 0.6360          & 0.2810          & 0.0000               & 0.0000               & 0.0078          \\
& Injection Attack~\cite{Revisiting}    & 0.6204          & 0.6243          & 0.3161          & 0.0000               & 0.0039          & 0.0312          \\
& \ourmethod{} (Ours)      & \textbf{0.7141}          & \textbf{0.7102}          & \textbf{0.3473}          & \textbf{0.0117}          & \textbf{0.0234}          & \textbf{0.3278}          \\
\midrule
\midrule
\multirow{2}{*}{\begin{tabular}[c]{@{}c@{}}Poisoning\\ Ratio ↓\end{tabular}} & Popularity→ & \multicolumn{3}{c|}{Lower Torso}                     & \multicolumn{3}{c}{Tail}                            \\ 
\cmidrule{2-8} 
& Model →     & WRMF            & ItemAE          & Mult-VAE        & WRMF            & ItemAE          & Mult-VAE        \\
\midrule
\multirow{3}{*}{0.1\%}                                                       
& Clean       & 0.0000          & 0.0000          & 0.0078          & 0.0000          & 0.0000          & 0.0195          \\
& Injection Attack~\cite{Revisiting}    & 0.0000          & 0.0000          & 0.0078          & 0.0000          & 0.0000          & 0.0273          \\
& \ourmethod{} (Ours)      & \textbf{0.0897}          & \textbf{0.0976}          & \textbf{0.0819}          & \textbf{0.1483}          & \textbf{0.1132}          & \textbf{0.0741}          \\
\midrule
\multirow{3}{*}{0.05\%}                                                      
& Clean       & 0.0000          & 0.0000          & 0.0234          & 0.0000          & 0.0000          & 0.0273          \\
& Injection Attack~\cite{Revisiting}    & 0.0000          & 0.0000          & 0.0117          & 0.0000          & 0.0000          & 0.0234          \\
& \ourmethod{} (Ours)      & \textbf{0.0156}          & \textbf{0.0039}          & \textbf{0.0546}          & \textbf{0.0039}          & \textbf{0.0039}          & \textbf{0.0585}          \\
\bottomrule
\end{tabular}
}
\end{table*}

\begin{revblock}
For quantitative comparison, we compare IndirectAD with the injection attack proposed by Tang et al.~\cite{Revisiting}. We select this method as our main baseline because it matches our evaluation protocol: offline fake-profile injection, target-item promotion, black-box victim recommenders, implicit-feedback top-\(K\) recommendation, and low poisoning ratios. It also generates fake user profiles to promote a target item and can be evaluated using the same datasets, victim models, poisoning ratios, and metrics as IndirectAD.
\end{revblock}

Additionally, we explored the transferability of the attack, utilizing a public dataset to generate fake users for injection into the recommender system. Given that this dataset is somewhat outdated, being collected a few years prior, we conducted the attack using half of the dataset and assessed the model trained on the entire dataset to determine whether the fake data, generated using a subset of the data, could be effectively transferred to a more current dataset incorporating additional data over time.

\subsection{Effectiveness of \ourmethod{}}
\label{subsec:Effectiveness}

We perform experiments across three recommender systems—WRMF, ItemAE, and Mult-VAE—utilizing two distinct poisoning ratios, and test these setups on four varied datasets: Books, Arts, Steam, and ML-1M.

\autoref{tab:main} shows the HR@20 of the target item across various configurations. We compare IndirectAD with the injection attack proposed by Tang et al.~\cite{Revisiting}, which is evaluated under the same offline fake-profile injection setting, datasets, victim models, poisoning ratios, and HR@20 metric. We note that, except when the dataset is ML-1M and the target model is ItemAE, IndirectAD consistently achieves higher HR@20 than this baseline under the low-poisoning-ratio setting. Generally, IndirectAD effectively boosts the target item even with a very low poisoning ratio. For instance, with the Amazon Books dataset and WRMF as the target model, the baseline injection attack fails at a poisoning ratio of only 0.05\%, showing \textit{zero} HR@20. In contrast, IndirectAD achieves an HR@20 of 0.3810\%. \rev{This result shows that indirect promotion can remain effective under a small attack budget on public benchmark datasets.} Furthermore, at a higher poisoning ratio (e.g., 0.1\%), IndirectAD continues to outperform the same baseline. For example, with the Amazon Books dataset and WRMF, the baseline method still shows \textit{zero} HR@20, whereas IndirectAD achieves an HR@20 of 1.1429\%. \rev{We interpret these results as evidence that low-ratio poisoning deserves careful consideration, while broader comparisons with other attack families depend on their threat models and evaluation protocols.}

Our method also exhibits strong transferability. When using WRMF as the substitute model, the attack successfully transfers to both ItemAE and Mult-VAE without significant loss of effectiveness. In some cases, performance on ItemAE and Mult-VAE even surpasses that on WRMF. For instance, with the Amazon Arts dataset and a 0.05\% poisoning ratio, our method achieves a 0.0285\% promotion on WRMF and 0.1680\% on Mult-VAE, indicating superior transferability compared to traditional injection attack techniques. \rev{The transferability mainly comes from the data-level nature of IndirectAD. The generated fake profiles alter the user-item interaction distribution by strengthening the association among the target-user group, the trigger item, and the target item. Although WRMF, ItemAE, and Mult-VAE use different model architectures, they all learn from the same interaction and co-occurrence signals. Therefore, poisoned profiles generated using the substitute model can still affect victim models with different architectures.}

\begin{table*}[!t]
\centering
\caption{The impact of the target user's diversity, measured by HR@20 (\%).}
\small
\label{tab:cluster}
\setlength{\tabcolsep}{3.5mm}{
\renewcommand{\arraystretch}{1.5}
\begin{tabular}{c|c|cc|cc|cc}
\toprule
\multirow{2}{*}{\begin{tabular}[c]{@{}c@{}}Poisoning\\ Ratio ↓\end{tabular}} & Model → & \multicolumn{2}{c|}{WRMF}          & \multicolumn{2}{c|}{ItemAE}        & \multicolumn{2}{c}{Mult-VAE}      \\ 
\cmidrule{2-8}
& Method ↓   & Clustering           & Random          & Clustering           & Random          & Clustering           & Random          \\
\midrule
\multirow{3}{*}{0.1\%}                                                       
& Clean     & 0.0000 & 0.0000 & 0.0000 & 0.0000 & 0.0000 & 0.0273 \\
& Injection Attack~\cite{Revisiting}  & 0.0000 & 0.0000 & 0.0000 & 0.0000 & 0.0952 & 0.0156 \\
& \ourmethod{} (Ours)    & \textbf{1.1429} & \textbf{0.4019} & \textbf{0.7619} & \textbf{0.2849} & \textbf{0.7619} & \textbf{0.4956} \\
\midrule
\multirow{3}{*}{0.05\%}                                                      
& Clean     & 0.0000 & 0.0000 & 0.0000 & 0.0000 & 0.0000 & 0.0078 \\
& Injection Attack~\cite{Revisiting}  & 0.0000 & 0.0000 & 0.0000 & 0.0039 & 0.0000 & 0.0312 \\
& \ourmethod{} (Ours)    & \textbf{0.3810} & \textbf{0.0117} & \textbf{0.0952} & \textbf{0.0234} & \textbf{0.4762} & \textbf{0.3278} \\
\bottomrule
\end{tabular}
}
\end{table*}

\subsection{Impact of Target Item Popularity}
\para{Setup}. This section examines the influence of item popularity on our method. We categorize item popularity into four levels: (i) head: items in the top 5\% of clicks; (ii) upper torso: items in the 75th to 95th percentile of clicks; (iii) lower torso: items in the 50th to 75th percentile of clicks; (iv) tail: items below the 50th percentile in clicks. We conduct this experiment using the Amazon Books dataset.

\begin{revblock}
\para{Observations}. \autoref{tab:pop} displays the effectiveness of IndirectAD across target items with different popularity levels. Overall, IndirectAD achieves higher HR@20 than the injection baseline~\cite{Revisiting} across all evaluated popularity groups under the same poisoning ratios and victim models.
When the poisoning ratio is 0.1\%, even for head items that already have relatively high visibility, IndirectAD increases the HR@20 of the target item from 0.5385\% to 0.9326\%, yielding an absolute improvement of 0.3941\%. In comparison, the injection baseline increases HR@20 to 0.6087\%, corresponding to an absolute improvement of 0.0702\%. Thus, under this setting, IndirectAD provides a stronger promotion effect than the baseline by leveraging the trigger item as an intermediate bridge.

Notably, the advantage of IndirectAD is also observed for less popular items. At the tail level with a poisoning ratio of 0.05\%, where the injection baseline provides little or no promotion effect, IndirectAD still achieves non-zero HR@20 across all evaluated victim models. This suggests that indirect promotion can help promote target items with different popularity levels under a low poisoning budget on public benchmark datasets.

It is important to note that item popularity is measured over all users, rather than only over the selected target-user group. Therefore, the HR@20 values do not necessarily follow the global popularity order of head, upper-torso, lower-torso, and tail items. This is because an item's overall popularity may differ from its relevance to the specific target-user group considered in the attack.
\end{revblock}

\subsection{Impact of Target User Diversity}
\para{Setup}. This section delves into how the diversity of the target user group affects our method. We consider two scenarios: (i) Clustering, where the attacker aims to promote the target item to a user group with distinct characteristics or similarities; and (ii) Random, where the target item is promoted to a randomly selected group of users. This experiment is conducted using the Amazon Books dataset, with the target item's popularity fixed at the upper torso level.

\begin{revblock}
\para{Observations}. \autoref{tab:cluster} illustrates the effectiveness of IndirectAD under both target-user selection strategies. Overall, IndirectAD achieves higher HR@20 than the injection baseline~\cite{Revisiting} in both clustered and random target-user settings. The advantage is more pronounced in the clustered setting, where target users share more similar interaction patterns, and is more modest in the random setting, where user preferences are more diverse. For instance, in the random setting with WRMF as the target model and a 0.1\% poisoning ratio, the injection baseline achieves zero HR@20, whereas IndirectAD achieves an HR@20 of 0.4019\%. In the clustered setting, IndirectAD also achieves substantially higher HR@20 than the baseline across the evaluated victim models and poisoning ratios. This suggests that the trigger item is more effective when the target-user group has coherent preferences, because a shared preference structure makes it easier to identify a trigger item that can serve as an intermediate bridge to the target item.

IndirectAD is less effective in the random setting than in the clustered setting. This is expected because a randomly selected target-user group may not share a clear preference pattern, making it more difficult to find a universally appealing trigger item. Nevertheless, IndirectAD still improves over the injection baseline in most evaluated random settings, indicating that indirect promotion can provide benefits even when the target-user group is more heterogeneous, although the attack effect is weaker than in the clustered setting.
\end{revblock}

\begin{table}[!t]
\centering
\caption{The impact of the trigger item selection, measured by HR@20 (\%).}
\small
\label{tab:trig}
\setlength{\tabcolsep}{1.2mm}{
\renewcommand{\arraystretch}{1.5}
\begin{tabular}{ccccc}
\toprule
\multirow{2}{*}{\begin{tabular}[c]{@{}c@{}}Poisoning\\ Ratio ↓\end{tabular}} & \multirow{2}{*}{Method ↓}                                  & \multirow{2}{*}{WRMF} & \multirow{2}{*}{ItemAE} & \multirow{2}{*}{Mult-VAE} \\
&                                                            &                       &                         &                           \\
\midrule
\multirow{4}{*}{0.1\%}                                                       & Clean                                                      & 0.0000                & 0.0000                  & 0.0000                    \\
& Injection Attack~\cite{Revisiting} & 0.0000                & 0.0000                  & 0.0952                    \\
& Popularity                             & 0.5268                & 0.4975                  & 0.2575                    \\
& \ourmethod{} (Ours)                             & \textbf{1.1429}       & \textbf{0.7619}         & \textbf{0.7619}           \\
\midrule
\multirow{4}{*}{0.05\%}                                                      & Clean                                                      & 0.0000                & 0.0000                  & 0.0000                    \\
& Injection Attack~\cite{Revisiting} & 0.0000                & 0.0000                  & 0.0000                    \\
& Popularity                             & 0.0976                & 0.0507                  & 0.1658                    \\
& \ourmethod{} (Ours)                             & \textbf{0.3810}       & \textbf{0.0952}         & \textbf{0.4762}     \\
\bottomrule
\end{tabular}
}
\end{table}

\begin{table*}[!t]
\centering
\caption{The impact of display window.}
\small
\label{tab:window}
\setlength{\tabcolsep}{1.5mm}{
\renewcommand{\arraystretch}{1.5}
\begin{tabular}{c|c|ccc|ccc|ccc}
\toprule
\multirow{2}{*}{\begin{tabular}[c]{@{}c@{}}Poisoning\\Ratio ↓\end{tabular}} 
  & \multirow{2}{*}{Method ↓} 
    & \multicolumn{3}{c|}{HR@10 (\%)} 
    & \multicolumn{3}{c|}{HR@20 (\%)} 
    & \multicolumn{3}{c}{HR@50 (\%)} \\
\cmidrule{3-11}
& & WRMF & ItemAE & Mult‑VAE & WRMF & ItemAE & Mult‑VAE & WRMF & ItemAE & Mult‑VAE \\
\midrule
\multirow{3}{*}{0.1\%}
  & Clean                                   & 0.0055 & 0.0000 & 0.0000 & 0.0000 & 0.0000 & 0.0273 & 0.0273 & 0.0109 & 0.0874 \\
  & Injection Attack~\cite{Revisiting}     & 0.0000 & 0.0000 & 0.0000 & 0.0000 & 0.0000 & 0.0156 & 0.0273 & 0.0109 & 0.0710 \\
  & \ourmethod{} (Ours)                     & \textbf{0.1857} & \textbf{0.1748} & \textbf{0.1803} 
                                           & \textbf{0.4019} & \textbf{0.2849} & \textbf{0.4956} 
                                           & \textbf{0.8140} & \textbf{0.6774} & \textbf{0.7812} \\
\midrule
\multirow{3}{*}{0.05\%}
  & Clean                                   & 0.0000 & 0.0000 & 0.0055 & 0.0000 & 0.0000 & 0.0078 & 0.0273 & 0.0055 & 0.1311 \\
  & Injection Attack~\cite{Revisiting}     & 0.0000 & 0.0000 & 0.0109 & 0.0000 & 0.0039 & 0.0312 & 0.0273 & 0.0164 & 0.1639 \\
  & \ourmethod{} (Ours)                     & \textbf{0.0055} & \textbf{0.0055} & \textbf{0.1311}
                                           & \textbf{0.0117} & \textbf{0.0234} & \textbf{0.3278}
                                           & \textbf{0.2458} & \textbf{0.2021} & \textbf{0.6119} \\
\bottomrule
\end{tabular}
}
\end{table*}

\begin{figure*}[!t]
\centerline{\includegraphics[width=\linewidth]{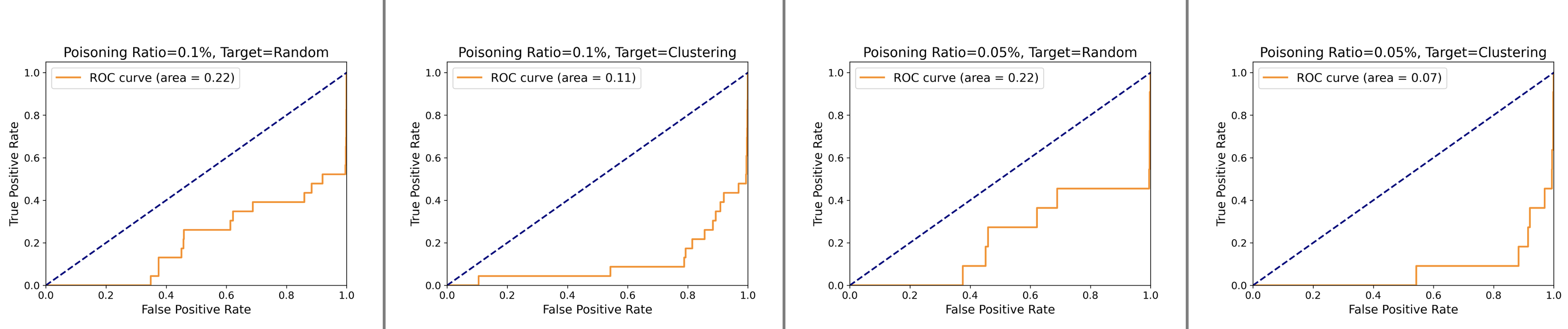}}
\caption{Detectability of \ourmethod{}.}
\label{fig:detection}
\end{figure*}

\begin{table*}[t]
\centering
\begin{revblock}
\caption{Impact of candidate-pool size on trigger-selection cost and attack effectiveness.}
\label{tab:candidate_pool}
\small
\begin{tabular}{lrrrrr}
\toprule
Candidate pool & \#Candidates & Selection time & HR@20(\%) \(\uparrow\) & NDCG@20(\%) \(\uparrow\) & Avg. rank \(\downarrow\) \\
\midrule
Top-50 popular items & 50 & 251.45s & \(0.3282\pm0.0264\) & \(0.0857\pm0.0073\) & \(20373\pm21.09\) \\
Top-100 popular items & 100 & 274.48s & \(0.3425\pm0.0052\) & \(0.0914\pm0.0014\) & \(20349\pm10.54\) \\
Top-500 popular items & 500 & 496.10s & \(0.3445\pm0.0032\) & \(0.0924\pm0.0069\) & \(20337\pm10.24\) \\
All post-filtered items & 20679 & 12238.72s & \(0.3608\pm0.0066\) & \(0.0955\pm0.0018\) & \(20338\pm24.56\) \\
\bottomrule
\end{tabular}
\end{revblock}
\end{table*}

\subsection{Impact of Trigger Item Selection}

\para{Setup.}  
Recall that by default, we select the trigger item based on its potential to be promoted using an optimization-based evaluation. Here, we explore how different selection methods for the trigger item may affect our attack. Specifically, we consider another intuitive trigger item selection approach based on popularity analysis, which we denote as “popularity.” It selects the trigger item by comparing the recommendation ranks given by the target user group \((U_t)\) versus the complement user group \((\overline{U_t})\), using the formula:

\[
I_{trig} = \argmin_i \bigl(Rank(i, U_t) - Rank(i, \overline{U_t})\bigr), \quad i \in I
\]

where \(I\) is the set of all items.  
We conduct this experiment on the Amazon Books dataset, with the target item’s popularity fixed at the upper-torso level.

\para{Observations.}  
\autoref{tab:trig} illustrates the effectiveness of our attack compared to the “popularity” baseline. We observe that our \ourmethod{} consistently outperforms the other approaches. This is likely because the “popularity” assessment relies on an item’s current state rather than its potential. In contrast, \ourmethod{} typically works by transforming the hard target into an easier one (the trigger), emphasizing an item’s potential rather than its current performance.

Although the “popularity” baseline performs worse than \ourmethod{}, it often outperforms the injection baseline~\cite{Revisiting}. This demonstrates that the high-level idea of indirect adversarial attacks provides stable and substantial benefits, even when using a suboptimal trigger item.

\rev{\autoref{tab:candidate_pool} shows that restricting the trigger candidate pool greatly reduces selection cost while preserving most of the attack effect. For example, the Top-500 pool reduces the selection time from 12238.72s to 496.10s, while retaining 95.5\% of the HR@20 and 96.8\% of the NDCG@20 achieved by searching all post-filtered items.}

\subsection{Impact of Display Window}
In prior sections we focused on HR@20—i.e.\ the hit ratio when recommending 20 items—since 20 is a realistic list length on most platforms. Here, we extend our evaluation to display windows of size 10, 20, and 50. We fix the dataset to Amazon Books and the target item's popularity to the upper‑torso level.

\autoref{tab:window} reports attack performance as the window size varies and for two poisoning ratios (0.1\% and 0.05\%). As expected, all methods see higher absolute HR as the window grows, but the gap between \ourmethod{} and the strongest baseline (the injection attack of \cite{Revisiting}) widens markedly. 

These results confirm two key points. First, larger display windows provide more ``slots'' for a Trojaned item to surface, boosting attack success for all methods.
Moreover, IndirectAD consistently outperforms the closest comparable injection baseline across the evaluated display windows and poisoning budgets, suggesting that the indirect-promotion mechanism remains effective under different top-\(K\) exposure settings.

\subsection{\rev{Detectability Analysis}}
\label{subsec:detect}
In this section, we investigate whether our attack can be easily detected by existing methods. We focus on the ``Catch the Black Sheep'' method proposed by Zhang et al.~\cite{DBLP:conf/ijcai/ZhangT0LCM15}. This method provides a unified framework for detecting poisoning attacks in recommender systems, leveraging a graph-based representation of user-item interactions. It identifies initial suspicious actions and propagates their influence through the network using label propagation to iteratively refine suspiciousness scores. By combining multiple detection techniques, this approach enhances accuracy and robustness against various shilling attack strategies. This method is a representative approach for detecting shilling profiles in real datasets.

\autoref{fig:detection} illustrates the detectability of the proposed attacks. We observe that the AUC score achieved by the ``Catch the Black Sheep'' method is extremely low, with the highest value being 0.22. Under the default scoring direction of this detector, the method does not assign higher suspiciousness scores to the injected profiles in our setting.

\rev{In addition to existing shilling-profile detection, we evaluate a simple co-occurrence-based statistical detector that directly targets the structural artifact of IndirectAD. This detector first computes a global co-occurrence score for each item pair from the observed user-item interaction data. Each profile is assigned a binary label, where injected fake profiles are labeled as 1 and benign profiles are labeled as 0. To obtain the anomaly score of a profile, we enumerate all item pairs contained in that profile, look up their precomputed co-occurrence scores, and aggregate them as the profile-level detection score. This detector does not assume prior knowledge of the attack-specific trigger-target pair, consistent with a realistic defense setting where the defender aims to identify anomalous users or interactions. Finally, we compute the AUC by comparing the profile-level scores with the binary labels. We normalize the aggregated score by the number of item pairs in the profile.}
\revminor{We evaluate this statistical detector under both 0.05\% and 0.1\% poisoning ratios, and the results are shown in \autoref{tab:detectability_stat_10} and \autoref{tab:detectability_stat_5}. The AUC values are deterministic under this detector, so zero standard deviations are omitted in these tables. Note that because only 5 fake users are injected at the 0.05\% poisoning ratio, the realized co-occurrence ratios are discretized by rounding to an integer number of profiles.}
\begin{table}[t]
\centering
\begin{revblock}
\caption{\revminor{Detectability under a simple statistical detector at 0.1\% poisoning ratio.}}
\label{tab:detectability_stat_10}
\small
\resizebox{0.48\textwidth}{!}{
\begin{tabular}{lccc}
\toprule
Method & \begin{tabular}{c}Trigger-target\\co-occurrence\\in fake users\end{tabular} & HR@20(\%) \(\uparrow\) & AUC \(\downarrow\) \\
\midrule
Injection Attack~\cite{Revisiting} & N/A & \(0.3059\pm0.0299\) & 0.6022 \\
IndirectAD-original & 100\% & \(0.7375\pm0.0696\) & 0.7208 \\
IndirectAD-relaxed & 70\% & \(0.5518\pm0.0449\) & 0.6742 \\
IndirectAD-relaxed & 50\% & \(0.6119\pm0.1050\) & 0.7248 \\
IndirectAD-relaxed & 30\% & \(0.4043\pm0.1116\) & 0.6138 \\
\bottomrule
\end{tabular}
}
\parbox{0.48\textwidth}{
\footnotesize
\vspace{4pt}
\revminor{The detection experiment settings are Amazon Books dataset and WRMF as both the substitute and victim models, with randomly selected target users.}
}
\end{revblock}
\end{table}

\begin{table}[t]
\centering
\begin{revminorblock}
\caption{\revminor{Detectability under a simple statistical detector at 0.05\% poisoning ratio.}}
\label{tab:detectability_stat_5}
\small
\resizebox{0.48\textwidth}{!}{
\begin{tabular}{lccc}
\toprule
Method & \begin{tabular}{c}Trigger-target\\co-occurrence\\in fake users\end{tabular} & HR@20(\%) \(\uparrow\) & AUC \(\downarrow\) \\
\midrule
Injection Attack~\cite{Revisiting} & N/A & \(0.1530\pm0.0311\) & 0.5515 \\
IndirectAD-original & 100\% & \(0.1857\pm0.0488\) & 0.6686 \\
IndirectAD-relaxed & 70\% & \(0.2731\pm0.0511\) & 0.5587 \\
IndirectAD-relaxed & 50\% & \(0.1967\pm0.0652\) & 0.5917 \\
IndirectAD-relaxed & 30\% & \(0.2021\pm0.1076\) & 0.5917 \\
\bottomrule
\end{tabular}
}
\parbox{0.48\textwidth}{
\footnotesize
\vspace{4pt}
\revminor{The detection experiment settings are Amazon Books dataset and WRMF as both the substitute and victim models, with randomly selected target users.}
}
\end{revminorblock}
\end{table}

\rev{\revminor{At the 0.1\% poisoning ratio,} the deterministic 100\% co-occurrence design achieves the strongest attack effect, but it is also more detectable under this statistical detector. 
Relaxed co-occurrence can reduce detectability when the co-occurrence ratio is sufficiently small. In particular, the 30\% relaxed variant still improves HR@20 over the injection baseline while bringing the AUC close to the baseline level.
\revminor{At the lower 0.05\% poisoning ratio, attack effectiveness does not vary monotonically with the configured co-occurrence ratio, partly because the small attack budget corresponds to only a few injected profiles, making the realized co-occurrence ratios discrete. More importantly, the detector AUC is consistently lower at 0.05\% than at 0.1\% across all evaluated settings, indicating that a smaller attack budget generally reduces detectability under this detector.}
These results show an effectiveness-detectability trade-off rather than proving that IndirectAD is undetectable against all possible detectors.}

\begin{table}[t]
\centering
\begin{revblock}
\caption{Sensitivity to the balance parameter \(\alpha\).}
\label{tab:alpha_sensitivity}
\small
\resizebox{0.48\textwidth}{!}{
\begin{tabular}{rrrrr}
\toprule
\(\alpha\) & HR@20(\%) \(\uparrow\) & NDCG@20(\%) \(\uparrow\) & MRR@20(\%) \(\uparrow\) & Avg. rank \(\downarrow\) \\
\midrule
0.1 & \(1.0448\pm0.0181\) & \(0.3522\pm0.0068\) & \(0.1684\pm0.0009\) & \(19638\pm57.66\) \\
0.3 & \(0.9628\pm0.0057\) & \(0.3232\pm0.0022\) & \(0.1537\pm0.0019\) & \(19655\pm32.12\) \\
0.5 & \(0.9549\pm0.0054\) & \(0.3261\pm0.0031\) & \(0.1577\pm0.0028\) & \(19670\pm75.41\) \\
0.7 & \(0.9668\pm0.0092\) & \(0.3366\pm0.0029\) & \(0.1682\pm0.0016\) & \(19677\pm34.40\) \\
0.9 & \(0.9848\pm0.0087\) & \(0.3486\pm0.0037\) & \(0.1788\pm0.0026\) & \(19628\pm46.96\) \\
\bottomrule
\end{tabular}
}
\end{revblock}
\end{table}

\begin{revblock}
\subsection{Sensitivity to the Balance Parameter}
\label{subsec:alpha}
We further evaluate the sensitivity of IndirectAD to the balance parameter \(\alpha\) in the composite loss. Recall that smaller \(\alpha\) emphasizes trigger-item promotion, while larger \(\alpha\) emphasizes target-item promotion. We use Amazon Books with WRMF as both the substitute and victim model, upper-torso target items, and clustered target users.

\end{revblock}

\rev{\autoref{tab:alpha_sensitivity} shows that the attack remains effective across a broad range of \(\alpha\) values. Although \(\alpha=0.5\) is not always the best-performing value, it provides a balanced default without tuning on the victim model or relying on additional target-system information.}

\section{Related Work}
\label{sec:related}
% \subsection{Trojan Attacks}
% A Trojan attack causes a victim DNN to output an adversary-chosen target label when the input contains a hidden trigger, posing risks to the DNN supply chain~\cite{backdoor, Blind, Trojaning, Hardware}. We focus on data-poisoning Trojan attacks, where the attacker cannot control the training process but instead injects trigger-embedded samples into the training data. Most prior poisoning-based Trojan attacks target image recognition~\cite{blended, lf, Input-Aware, SSBA, WaNet}; a representative example is BadNets~\cite{backdoor}, which stamps a trigger on a small subset of training samples and relabels them to the target class. Although \ourmethod{} draws on the trigger-based intuition of Trojan attacks, it differs substantially because recommender systems and image classifiers rely on fundamentally different data structures and prediction mechanisms.

% \subsection{Attacks to Recommender Systems}
The first widely recognized method of poisoning attacks on recommender systems is known as the shilling attack. Initially proposed in \cite{shillinga}, with theoretical analysis provided, shilling attacks were further explored by Lam et al.~\cite{shilling}, who conducted systematic research on their effects on both user-based and item-based recommender systems. They optimized shilling attacks through the use of random and average attacks in the creation of fake user profiles. Mobasher et al.~\cite{shillingb} introduced advanced methods, including consistency, segment, and bandwagon attacks, capable of executing both push and nuke attacks to manipulate item popularity rankings. Furthermore, Mobasher et al.~\cite{shillingc} proposed the love/hate and reverse bandwagon attacks specifically for nuke attacks. Zhang et al.~\cite{shillinghybrid} and Su et al.~\cite{shillinggroup} expanded upon these methods with the introduction of hybrid and group attacks, respectively, the latter considering the relationships between fake user profiles during construction. Subsequent studies have largely focused on user-based systems, while Seminario et al.~\cite{PIAttack} provided effective attacks on item-based systems.

Recent advancements have incorporated machine learning techniques to enhance profile data poisoning attacks. Li et al.~\cite{datapoisonCF} delivered a theoretical analysis and implemented attacks in collaborative filtering systems through optimization problems, integrating modern machine learning approaches into shilling attacks. This was followed by Christakopoulou et al.~\cite{Adversarial} and Fang et al.~\cite{influence}, who applied optimization methods to their work. Lin et al.~\cite{augment} and Zhang et al.~\cite{RecUP} utilized Generative Adversarial Networks (GANs) to generate fake user profiles based on real user data, with a discriminator to differentiate between real and fake users.

As recommender system technologies, especially those incorporating machine learning, have evolved, so too have attacks on leading platforms and advanced recommendation techniques. Xing et al.~\cite{XSS} executed attacks on major platforms like YouTube, Google, and Amazon using XSRF to inject fake user clicks. Yang et al.~\cite{Yang2017FakeCI} transformed fake user profile generation into an optimization problem, attacking websites including YouTube, Amazon, eBay, and Yelp. Fang et al.~\cite{graphbase} and Thanh et al.~\cite{poisonGNN} targeted graph-based recommender systems with data poisoning attacks, while Yue et al.~\cite{sequence} focused on sequence-model based systems. Song et al.~\cite{poisonrec} applied reinforcement learning and sequence model neural networks for black-box attacks on multiple recommendation models, with Zhang et al.~\cite{ReverseAttack} and Chen et al.~\cite{KGAttack} utilizing surrogate models and knowledge graphs, respectively, to facilitate such attacks.

\rev{These prior attacks are closely related but differ from IndirectAD in their attack interfaces and objectives. Some methods formulate poisoning as direct target-item promotion, some rely on GAN-based profile generation, some exploit platform-specific co-visitation signals, and others assume sequential or interactive attack processes. In contrast, IndirectAD uses a trigger item as an intermediate bridge and injects low-ratio fake profiles into an offline user-item interaction matrix. This design targets the setting of black-box victim recommenders and implicit-feedback top-\(K\) recommendation, which is why we use the closest comparable fake-profile injection attack as the main quantitative baseline.}

\section{Discussion}
\label{sec:discussion}

\begin{revblock}
% \para{Interpreting the attack effect.}
% The HR@20 values achieved by \ourmethod{} may appear numerically modest in some settings, but they should be interpreted in the context of low-ratio item-promotion attacks. The attacker controls only a very small fraction of the user base, and the target item is evaluated under a top-20 exposure metric. In this setting, even a small increase in HR@20 indicates that the injected profiles can shift the ranking outcome for a measurable fraction of the target users. Our display-window analysis further shows that the attack effect becomes more visible when the recommendation window is enlarged, suggesting that \ourmethod{} affects not only whether the target item appears in the top-20 list but also its broader ranking position.

\para{Limitations and real-world constraints.}
Our evaluation demonstrates the effect of low-ratio poisoning on public real-world benchmark datasets, but it should not be interpreted as direct evidence of performance on proprietary industrial-scale platforms. Real-world deployment would face additional constraints, including fake account creation cost, temporal interaction consistency checks, bot detection, anomaly detection, and graph-based fraud detection. Similarly, our main threat model assumes access to same-domain public or partially available data for substitute-model training. A fully external proxy dataset may introduce a larger distribution gap and reduce attack transferability. Therefore, \ourmethod{} should be viewed as exposing a low-ratio data-poisoning risk under the evaluated threat model, rather than as an end-to-end attack against all production-level defenses.

\para{Defense implications.}
The detectability analysis suggests several possible defense directions. Since \ourmethod{} relies on strengthening associations between the trigger and target items, defenders may monitor sudden changes in pairwise item co-occurrence, lift, or PMI, especially for low-popularity target items. Profile-level anomaly detection can also examine whether a group of newly created profiles shares unusually similar item-pair patterns. More advanced defenses may combine statistical co-occurrence monitoring, temporal interaction modeling, graph-based fraud analysis, and robust recommender training. We do not claim robustness against all such production-level defenses; evaluating \ourmethod{} against these defense pipelines is a direction for future work.

\subsection{Potential Countermeasures}

\para{Detection.}
Detection-based defenses aim to identify suspicious profiles or abnormal interaction patterns before they substantially affect model training. Earlier methods used manually designed profile features and classical classifiers such as SVM or KNN to detect anomalous user profiles~\cite{Burke,ZhangZ14}. Unsupervised methods, including probabilistic latent semantic analysis and clustering-based approaches, have also been used to identify outlier profiles~\cite{Mehta,Bhaumik}. More recent defenses use sequential or graph-based models to capture richer behavioral signals. For example, LSTM-based methods can analyze user behavior sequences~\cite{Gao}, while GNN-based methods can jointly model user-item interactions and propagate suspiciousness over the interaction graph~\cite{Zhang20}. These methods are relevant to \ourmethod{} because the injected profiles may introduce unusual co-occurrence or graph patterns, especially when trigger-target co-occurrence is enforced too deterministically.

\para{Adversarial robust training.}
Another line of defense improves the robustness of recommender models during training. For example, Adversarial Personalized Ranking perturbs user and item embeddings to improve the robustness of BPR-based matrix factorization~\cite{He}. Follow-up work extends adversarial training to multimedia and context-aware recommendation settings~\cite{TangDHYTC20, ChenL19}. These methods are complementary to detection-based defenses: instead of identifying suspicious profiles directly, they aim to reduce the model's sensitivity to adversarially crafted interactions. Evaluating whether such robust training methods can mitigate low-ratio trigger-mediated poisoning remains an important direction for future work.
\end{revblock}

\section{Conclusions}
\label{sec:conclusion}

\begin{revblock}
In this study, we introduced \ourmethod{}, a low-ratio data poisoning attack for item promotion in recommender systems. \ourmethod{} uses a trigger item as an intermediate bridge between the target users and the target item, allowing the attacker to promote the target item with fewer injected fake profiles than direct-promotion baselines. Our evaluations on four public benchmark datasets and three recommender models show that \ourmethod{} can improve target-item exposure under poisoning ratios as low as 0.05\%. These results suggest that low-ratio poisoning remains a relevant security concern for recommender systems, even when the attacker controls only a small fraction of the user base. At the same time, our findings should be interpreted under the evaluated threat model and public benchmark settings; robustness against production-level fraud detection, bot detection, and other anti-abuse mechanisms remains an important direction for future work.
\end{revblock}
% \appendix
% \section{Location}

% \clearpage
\bibliographystyle{IEEEtran}
\bibliography{bib}

\end{document}